\documentclass[12pt]{article}

\usepackage{amsmath,amssymb}
\usepackage{graphicx}
\usepackage{caption}
\usepackage{color}
\usepackage{dcolumn}
\usepackage{bm}
\usepackage[numbers,super,comma,sort&compress]{natbib}

\definecolor{background-color}{gray}{0.98}

\title{The chemical physics of unconventional superconductivity}
\author{Sumit Mazumdar\thanks{Department of Physics, University of Arizona, Tucson, AZ 85721}, 
R. Torsten Clay\thanks{Department of Physics \& Astronomy and HPC$^2$
Center for Computational Science, Mississippi State University, 
Mississippi State, MS 39762}}

\begin{document}

\maketitle

\begin{abstract}
Attempts to explain correlated-electron superconductivity have largely
focused on the proximity of the superconducting state to
antiferromagnetism. Yet, there exist many correlated-electron systems
that exhibit insulator-superconducting transitions where the insulating
state exhibits spatial broken symmetry different from
antiferromagnetism. Here we focus on a subset of such compounds which
are seemingly very different in which specific chemical
stoichiometries play a distinct role, and small deviations from
stoichiometry can destroy superconductivity. These superconducting
materials share a unique carrier concentration, at which we show there
is a stronger than usual tendency to form local spin-singlets. We
posit that superconductivity is a consequence of these pseudomolecules
becoming mobile as was suggested by Schafroth a few years prior to
the advent of the BCS theory. 

\end{abstract}

\clearpage

  \makeatletter
  \renewcommand\@biblabel[1]{#1.}
  \makeatother

\renewcommand{\baselinestretch}{1.5}
\normalsize

\clearpage

\section*{\sffamily \Large INTRODUCTION}

Theoretical condensed matter physicists have been searching for a
theory of correlated-electron superconductivity (SC) for more than 25
years, since the discovery\cite{Bednorz86a} of SC in
La$_{2-x}$Sr$_x$CuO$_4$.  Consensus is slowly emerging that the
problem demands a conceptually new approach altogether. It is also
accepted by many scientists by now that copper oxides are but only one
out of many families or classes of materials in which SC is
unconventional, in the sense that the SC cannot be explained within
the standard BCS approach.  Materials in which SC is thought to be
unconventional include besides the cuprates the new Fe compounds
\cite{Wen11a}, various ternary and quaternary transition metal
compounds \cite{Johnston73a,Hagino95a}, organic charge-transfer solids
\cite{Ishiguro}, and perhaps also the fullerides \cite{Capone09a} and
the recently discovered metal-intercalated polycyclic aromatic
hydrocarbons \cite{Kubozono11a} such as phenanthrene, picene, etc. In
all these cases electron-electron (e-e) interactions are believed to
be strong. As shown by Uemura et al. two decades back
\cite{Uemura91a}, unconventional superconductors can be be identified
by their large T$_c$/T$_F$ (here T$_F$ is the Fermi temperature).
Thus while T$_c$/T$_F$ $\sim$ 10$^{-5}$ for elemental Al and Zn, and
$\sim$10$^{-3}$ for Nb with the highest T$_c$ among elements, the
unconventional superconductors all lie within a band with
10$^{-2}<$T$_c$/T$_F<$10$^{-1}$ (see Fig.~3 in reference \citenum{Uemura91a}).

While the bulk of the theoretical effort has gone into attempts to
understand the detailed behavior of individual families of materials
(such as the origin of the pseudogap in the cuprates), an alternate
approach involves determining {\it what} precisely is common between
these materials {\it besides} strong e-e interactions, because not all
strongly correlated systems are superconducting. It is here that we
believe that understanding of certain {\it chemical} features of the
unconventional superconductors gain relevance. In other words, we
believe that the physics of unconventional superconductors is very
strongly determined by their chemistry. This is the topic of this
Review. In the following we attempt to show that many
correlated-electron superconductors share two common features, (i)
carrier density $\rho$ of exactly $\frac{1}{2}$ per atom, molecule or
unit cell; and (ii) lattice frustration. Materials possessing these
two features exhibit a strong tendency to form local spin-singlets
that are the Bosonic pseudomolecules in Schafroth's theory of SC
\cite{Schafroth55a}.  In this Review we first discuss these features
in the context of the organic charge-transfer solids, and then show
that similarities can be found in several other seemingly unrelated
classes of unconventional superconductors.  We recognize that there
exist other correlated-electron superconductors that are not
$\rho=\frac{1}{2}$.  Even here we believe that formation of local
spin-singlets can occur and Schafroth's theory is relevant. For
example, in the context of cuprates many scientists hold the opinion
that preformed Cooper pairs form at temperatures much higher than
T$_c$ and condense only at T$_c$.  The actual demonstration of local
singlets in these other superconductors will require further work.

\section*{\sffamily \Large Charge-transfer Solids as prototype $\rho=\frac{1}{2}$ superconductors}

SC in organic charge-transfer solids (CTS) has been
known\cite{Jerome80b} since 1980.  The two most well known families of
superconducting CTS are the (TMTSF)$_2$X and (BEDT-TTF)$_2$X, where
the molecules TMTSF and BEDT-TTF constitute the active components
containing the charge-carrying holes and X are closed shell anions
\cite{Ishiguro}.  While conducting CTS compounds exist with range of
charge transfers $0.5 \leq \rho < 1$ between cations and anions, in
all cases the stoichiometry is 2:1 for the cationic superconductors
and 1:2 for anionic superconductors. Thus the carrier concentration
per {\it molecule}, which is how we define $\rho$ is invariably
$\frac{1}{2}$. We believe that requirement of a specific density for
SC here is an important feature.

\subsubsection*{\sffamily \normalsize Effective $\rho=1$ model}

The highest T$_c$ in the CTS is found in the $\kappa$-(BEDT-TTF)$_2$X,
in which there occur dimers of BEDT-TTF molecules, with strong
intradimer electron hoppings and weaker interdimer hoppings
\cite{Kanoda11a}. The dimers form anisotropic triangular lattices. At
ambient pressures and low temperatures, $\kappa$-(BEDT-TTF)$_2$X are
antiferromagnetic (AFM) insulators, and under moderate pressure they
become superconducting \cite{Kanoda11a}. The AFM is described easily
within an {\it effective} $\frac{1}{2}$-filled band ($\rho=1$) Hubbard
model (with each dimer an effective site) that is close to being a
square lattice, given by the following Hamiltonian:
\begin{equation}
H=-t\sum_{\left< ij\right>}B_{i,j}-t^\prime\sum_{\left[ ij\right]}B_{i,j}+
U\sum_i n_{i\uparrow}n_{i\downarrow}
\label{ham}
\end{equation}
In Eq.~\ref{ham}
$B_{i,j}=\sum_\sigma(c^\dagger_{j,\sigma}c_{i,\sigma}+H.c.)$ is the
kinetic energy operator for the bond between sites $i$ and $j$, where
$c^\dagger_{i,\sigma}$ creates an electron of spin $\sigma$ on site
$i$.  The sites $i$ and $j$ in $\left<ij\right>$ are nearest neighbors
on a square lattice while $\left[ij\right]$ are sites connected in the
$x+y$ direction (see Fig.~3(c) in reference \citenum{Kanoda11a}).
$n_{i\sigma}=c^\dagger_{i,\sigma}c_{i,\sigma}$ is the density operator
and $n_i=n_{i\uparrow}+n_{i\downarrow}$.  $U$ is the on-site Coulomb
interaction.

\begin{figure}
\begin{minipage}{0.4\columnwidth}
\centerline{\includegraphics[width=2.4in,keepaspectratio=true]{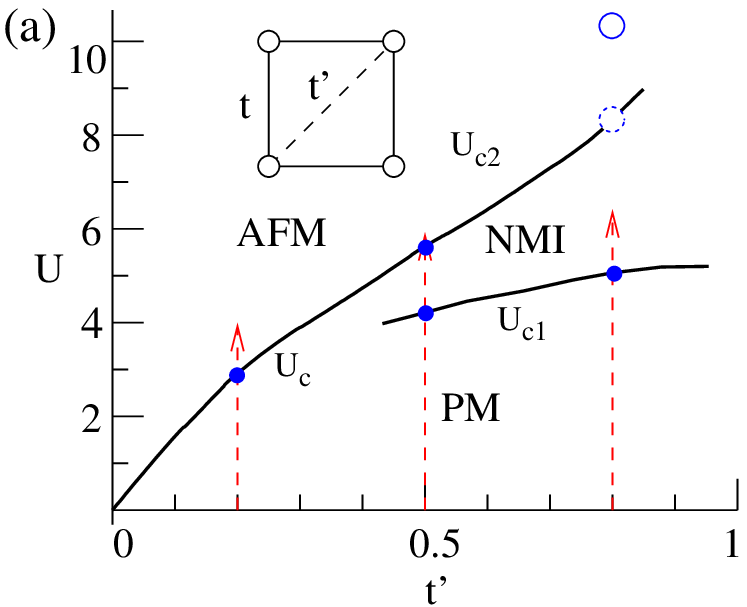}}
\end{minipage}
\begin{minipage}{0.5\columnwidth}
\centerline{\includegraphics[width=3.5in,keepaspectratio=true]{fig1bcde.eps}}
\end{minipage}
\caption{\label{rho1fig} (a) Phase diagram of the 2D $\rho=1$ effective 
model of Eq.~\ref{ham} based on PIRG calculations \cite{Dayal12a}.
Filled points are determined using PIRG and finite-size scaling. The
NMI/AFM phase boundary at $t^\prime$=0.8 is more uncertain; the
solid circle there is the upper bound from 4$\times$4 exact
diagonalization, and the dotted circle is the expected boundary
in the thermodynamic limit. (b)-(d) Double occupancy, 
$t^\prime$ bond order, and long-distance $d_{x^2-y^2}$ pair-pair correlation
function, respectively, as a function of $U$ for $t^\prime$=0.5.
Squares (diamonds) are for 6$\times$6 (8$\times$8) lattices.
(e) Enhancement of the pair-pair correlation over the uncorrelated
system. Here circles are exact results for the 4$\times$4 lattice.
Reproduced from Reference \citenum{Dayal12a}.}
\end{figure}

The ground state of Eq.~\ref{ham} in the $t^\prime=0$ limit is the
Ne\'el AFM state.  This had prompted some scientists to propose that
pressure reduces the lattice anisotropy (increasing the isotropic
character) and increases the bandwidth, and at a critical bandwidth SC
dominates over AFM.  The phase diagram of Eq.~\ref{ham} as determined
\cite{Morita02a,Dayal12a} using the using the Path Integral
Renormalization Group (PIRG) method \cite{Kashima01b} is shown in
Fig.~\ref{rho1fig}(a).  As the frustration $t^\prime$ increases from
zero, a paramagnetic metallic (PM) enters. The metal-insulator
transition here may be seen in a simultaneous drop in the double
occupancy ($D=\langle n_{i,\uparrow}n_{i,\downarrow}\rangle$) and the
bond order ($B=\sum_{\sigma}\langle
c^\dagger_{i,\sigma}c_{j,\sigma}\rangle$) as $U$ increases at fixed
$t^\prime$ (see Fig.~\ref{rho1fig}(b) and (c)).  At still larger
$t^\prime$, a non-magnetic insulator (NMI) phase which unlike the AFM
phase has no long-range magnetic order
\cite{Kashima01a,Morita02a,Mizusaki06a}, enters between the PM and AFM
phases.

Many mean-field calculations suggested that SC occurs near the
metal-insulator transition in the model (see Reference
\citenum{Dayal12a} for a discussion of these papers).  As we have
investigated Eq.~\ref{ham} with a fixed number of particles, we looked
for off-diagonal long range order (ODLRO) \cite{Yang62a} by
numerically calculating the pair-pair correlation function.  The
operator $\Delta^\dagger_{i,j}$ creates a singlet pair on lattice
sites $i$ and $j$:
\begin{equation}
\Delta^\dagger_{i,j}= \frac{1}{\sqrt{2}}
(c^\dagger_{i,\uparrow}c^\dagger_{j,\downarrow}
- c^\dagger_{i,\downarrow}c^\dagger_{j,\uparrow}).
\end{equation}
The pair-pair correlation function is then defined as
\begin{equation}
P({\bf r})=\frac{1}{4}\sum_\nu g(\nu)\langle \Delta^\dagger_i
\Delta_{i+\bf{r}(\nu)}\rangle.
\label{pair}
\end{equation}
In Eq.~\ref{pair} the sum is over the four nearest neighbor sites of
the square lattice; the phase factor $g(\nu)$ determines the symmetry
of the superconducting order parameter.  We have performed explicit
calculations of P({\bf r}) for $s$ ($g(\nu)=1$ for all $\nu$) and
$d_{x^2-y^2}$ ($g(\nu)$ alternating $\pm$ 1) pair symmetries within
Eq.~\ref{ham}.  If SC is present, $P({\bf r})$ measured in the ground
state must converge to a nonzero value for $|{\bf
  r}|\rightarrow\infty$.  One also expects an enhancement of $P(r)$ by
the $U$ interaction.  In calculations of $P({\bf r})$ using exact
diagonalization \cite{Clay08a} and on larger lattices using PIRG
\cite{Dayal12a}, $P({\bf r})$ for all ${\bf r}$ beyond nearest
neighbor pair separation decreases continuously with increasing $U$
(see Fig.~\ref{rho1fig}(d)), consistent with the absence of SC in the
model. This is shown in Fig.~\ref{rho1fig}(d), where we plot the
pair-pair correlation $P_d(r^\star)$ for $d_{x^2-y^2}$ symmetry, where
$r^\star$ corresponds to one of the longest pair separations possible
on each finite lattice \cite{Dayal12a}.  In Fig.~\ref{rho1fig}(e) we
plot the {\it difference} $\Delta P_d(r,U)=P_d(r,U)-P_d(r,U=0)$
showing the enhancement of the pairing over the uncorrelated model; we
find no enhancement beyond nearest-neighbor distances. The small
enhancement for nearest-neighbor pairs (which overlap in real space)
is likely the reason that mean-field methods find SC in the model
\cite{Dayal12a}.

Other numerical studies going beyond the mean-field level also fail to
find SC \cite{Tocchio09a}. More recently, we have shown that the
addition of an additional AFM Heisenberg interaction $J_{ij}$ to
Eq.~\ref{ham} also fails to produce SC \cite{Gomes13a}.
Experimentally, the insulating phase proximate to SC in the CTS can be
different from AFM, including charge-ordered CO or the so-called
valence-bond-solid phase. Neither of these insulating states are
accounted for within the effective $\rho=1$ model \cite{Gomes13a}.

\subsubsection*{\sffamily \normalsize $\rho=\frac{1}{2}$ model and quasi-1D CTS}

An alternate approach to the effective $\rho=1$ model is the
$\rho=\frac{1}{2}$ model where we consider individual molecules and
not dimers as the proper units. We have done calculations within the
extended Hubbard model both in one and two dimensions (1D and 2D) that
show the strong tendency to form nearest neighbor spin-singlets in
this case.  When a nearest neighbor singlet forms between two
molecules in a system with an average charge $\rho=\frac{1}{2}$,
necessarily the charge density on the molecules involved in the bond
is slightly increased, $\rho_+=0.5+\delta$, while the charge density
on the non-bonded molecules is slightly decreased,
$\rho_-=0.5-\delta$. Thus the formation of singlet pairs in a
$\rho=\frac{1}{2}$ system implies the presence of charge-ordering or
at least charge-disproportionated molecules.

 The general form of the Hamiltonian we consider 
for these systems is the following Peierls Extended Hubbard model:
\begin{eqnarray}
H&=&-\sum_{\langle ij\rangle}t_{ij}(1+\alpha\delta_{i,j})B_{i,j}
 +\frac{1}{2}K_\alpha\sum_{\langle ij\rangle}\delta^2_{i,j} + \beta\sum_i n_i v_i \nonumber \\
&+& \frac{1}{2}K_\beta \sum_i v_i^2 +  U \sum_i  n_{i\uparrow}n_{i\downarrow} + 
V \sum_{\langle ij\rangle}  n_in_j.
\label{ham1d}
\end{eqnarray}
The terminology in Eq.~\ref{ham1d} follows that of Eq.~\ref{ham}.  In
addition to the onsite Coulomb interaction $U$, we include in general
the nearest-neighbor Coulomb interaction $V$.  Electron-phonon (e-p)
coupling is included in the semi-classical approximation, where
$\alpha$ ($\beta$) is the inter-site (intra-site) e-p coupling
constant and $K_\alpha$ ($K_\beta$) the associated spring constant.
We solve Eq.~\ref{ham1d} numerically, measuring the charge density
$\langle n_i\rangle$ and bond order $\langle B_{i,j}\rangle$.  The
classical inter-- and intra--molecular distortions $\delta_{i,j}$ and
$v_i$ are determined self-consistently \cite{Clay03a} from the
equations
\begin{equation}
\delta_{i,j}=-\frac{\alpha}{K_\alpha}\langle B_{i,j}\rangle,
\qquad v_i=-\frac{\beta}{K_\beta}\langle n_i\rangle.
\end{equation}
Other correlation functions such as spin-spin correlations,
$\langle S^z_i S^z_j\rangle=\langle (n_{i,\uparrow}-n_{i,\downarrow})(n_{j,\uparrow}-n_{j,\downarrow})\rangle$,
may be measured following convergence of the iterative 
self-consistency procedure.

\begin{figure}
\centerline{\includegraphics[width=3.0in,keepaspectratio=true]{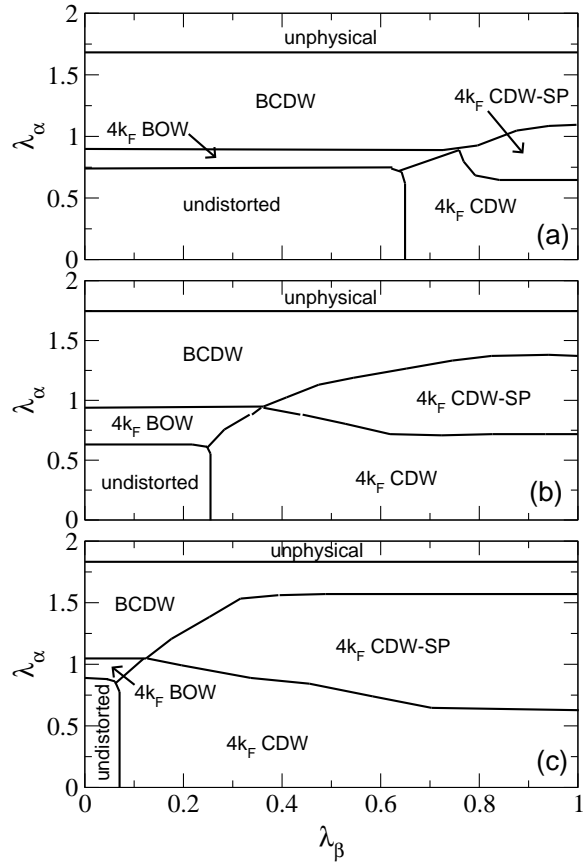}}
\caption{\label{1dfig}Phase diagram of the 1D model of
Eq.~\ref{ham1d} based on 16-site exact calculations for $U=8$ and  (a) $V=2$, (b)
$V=3$, and (c)  $V=4$. $\lambda_\alpha$ and $\lambda_\beta$ are
the normalized inter- and intra-site e-p couplings (see text).
Reproduced from Reference \citenum{Clay03a}.}
\end{figure}

The ground state of Eq.~\ref{ham1d} is well understood in the 1D limit
where a number of different broken-symmetry phases are found.  In
Fig.~\ref{1dfig} we show the phase diagram of Eq.~\ref{ham1d} for a 1D
16 site lattice, with e-e parameters chosen as appropriate for the
(TMTTF)$_2$X group of materials \cite{Clay03a}.  The phase diagrams
are plotted as a function of the normalized e-p couplings constants
$\lambda_\alpha=\alpha^2/(K_\alpha t)$ and
$\lambda_\beta=\beta^2/(K_\beta t)$. At $\rho=\frac{1}{2}$ there is a
competition between two different insulating phases: First, the
nearest-neighbor Coulomb interaction $V$ in Eq.~\ref{ham1d} leads to a
charge-ordered state (labeled ``4k$_{\rm F}$ CDW'' in
Fig.~\ref{1dfig}) with alternating charge densities
large--small--large--small in the pattern ``1010''.  In the 1D system
this state occurs for $V>V_c$, where $V_c=2t$ for
$U\rightarrow\infty$, and $V_c>2t$ for finite $U$.  Sufficiently
strong e-p coupling can lead to a spin-Peierls (SP) state (4k$_{\rm
  F}$ CDW-SP in Fig.~\ref{1dfig}), where the spin-singlet bonds
between the charge-rich sites alternate in strength (i.e.,
bond-distortion pattern ``strong-strong-weak-weak'',
$[1=0=1\;\mbox{--}\;0\;\mbox{--}\;]$, where a ``double'' bond is
stronger than a ``single'' bond).  Secondly, for $V<V_c$, a
charge-ordered state with charge pattern 1100 is found. In this
Bond-Charge Density Wave (``BCDW'' in Fig.~\ref{1dfig}) state,
nearest-neighbor singlets form between the charge-rich sites and bond
orders are also necessarily modulated.  In the BCDW the bond pattern
may be either ``strong-undistorted-weak-undistorted'',
$[1=1\;\mbox{--}\;0\cdots0\;\mbox{--}]$, or
``strong-weak-strong-weak$^\prime$ '', $[1=0\cdots0=1\;\mbox{--}\;]$,
depending on the strength of e-e correlations \cite{Ung94a}.  The
singlet formation in the BCDW leads to a nonmagnetic ground state with
a spin gap.  The SP state that is observed experimentally in the
quasi-1D $\rho=\frac{1}{2}$ CTS in all cases is the BCDW and not the
4k$_{\rm F}$ CDW-SP.  Compared to the 4k$_{\rm F}$ CDW, the charge
density modulation in the BCDW is much smaller, and the clearest
experimental signatures are the presence of a spin gap and the
predicted bond distortion pattern.  In materials where the bond
pattern in the SP state has been measured, for example MEM(TCNQ)$_2$,
the measured bond pattern is the same as that predicted for the BCDW
from calculations \cite{Clay03a}.

\subsubsection*{\sffamily \normalsize AFM to PEC transition and 2D CTS}

\begin{figure}
\centerline{\includegraphics[width=3.5in,keepaspectratio=true]{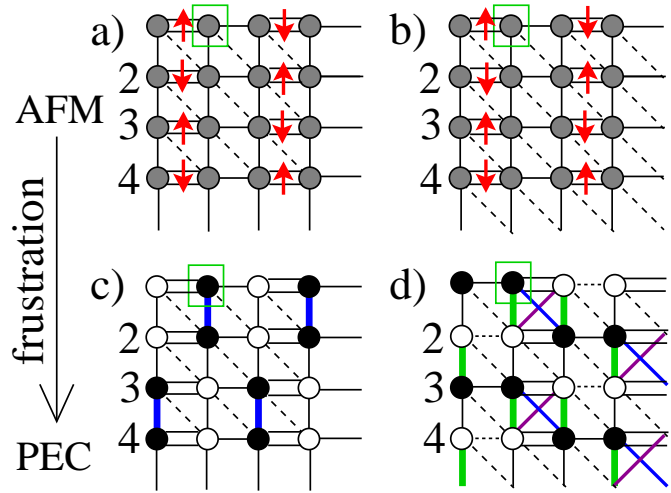}}
\caption{\label{pecfig}The AFM to PEC transition as seen in exact
  4$\times4$ cluster calculations for dimerized $\rho=0.5$
  systems. Panels (a) and (c) correspond to calculations in open
  boundary conditions, while those in (b) and (d) are from periodic
  boundary calculations. Grey circles correspond to sites with charge
  density $\rho=0.5$; black (white) circles to sites with $\rho>0.5$
  ($\rho<0.5$). The arrows in (a) and (b) indicate the observed AFM
  pattern observed in spin correlations. The heavy lines in (c) and
  (d) indicate the location of nearest-neighbor singlet bonds formed
  in the PEC state.  Reproduced from Reference \citenum{Li10a}.}
\end{figure}

In a 2D square lattice of dimers the ground state of Eq.~\ref{ham1d} for
finite $U$ and $V<V_c$ at $\rho=\frac{1}{2}$ has AFM order (see
Fig.~\ref{pecfig}(a) and (b)).  If sufficient lattice frustration is introduced
the AFM order is expected to vanish.  
The lattice structure we consider here is
a square lattice with dimerization along the $x$ direction. A
$t^\prime$ bond in the $x+y$ direction introduces frustration. 
To understand the effect of frustration we calculated charge
densities, bond orders, and most importantly spin-spin correlation
function as a function of $t^\prime$. 
Details of these results are shown
in original work \cite{Li10a}.  Above a critical value of $t^\prime$ a
sudden change occurs in all of these correlation functions. At this
transition, the AFM order shown in Fig.~\ref{pecfig}(a) and (b) gives
way to the charge-ordered state shown in Fig.~\ref{pecfig}(c) and (d).
Examination of the spin-spin correlations demonstrates the loss of AFM
order and formation of local singlets \cite{Li10a}.

We have termed this state a Paired Electron Crystal (PEC)
\cite{Li10a}. In the PEC the same local charge order (CO) pattern
$\cdots$1100$\cdots$ is found as in 1D. Schematic figures showing
these results are shown in Fig.~\ref{pecfig}, where
Fig.~\ref{pecfig}(c) shows the PEC state found under open boundary
conditions and Fig.~\ref{pecfig}(d) the PEC state found under periodic
boundary conditions. 

The PEC state has been seen experimentally in a number of 2D CTS.  One
example of a class of CTS that well illustrates the AFM/PEC
phenomenology as lattice frustration is varied is the
Z[Pd(dmit)$_2$]$_2$ series \cite{Tamura09a}.  Like the
$\kappa$-(BEDT-TTF)$_2$X, in the Z[Pd(dmit)$_2$]$_2$ crystal structure
Pd(dmit)$_2$ occur in dimers. Through different choices in the cation
Z, which change the crystal anisotropy, a series of ground states from
AFM to charge and ``valence bond'' order are seen \cite{Tamura09a}.
The transition between the Mott insulator with uniform dimer charges
and the PEC state has also been studied experimentally
\cite{Okazaki13a} in $\beta$-({\it meso}-DMBEDT-TTF)$_2$PF$_6$.
Certain CTS with the $\kappa$-(BEDT-TTF)$_2$X structure do show CO
states. For example, in
$\kappa$-(ET)$_4$[M(CN)$_6$][N(C$_2$H$_5$)$_4$]$\cdot$2H$_2$O (M= Co,
Fe), a transition from a Mott insulating phase to a CO
spin gap phase is found as the temperature goes below T=150K \cite{Ota07a}. Here and in many
other examples, evidence of fluctuating CO is found before
the transition (T$>$150 K) \cite{Ota07a}.  Further experimental
evidence for the PEC in 2D CTS is discussed in references
\citenum{Li10a} and \citenum{Dayal11a}.

\subsubsection*{\sffamily \normalsize Model for SC}

In many cases the experimentally seen PEC state is adjacent to SC. For
example, under ambient pressure, EtMe$_3$P[Pd(dmit)$_2$]$_2$ has an
insulating PEC ground state with inter-dimer singlet pairs (termed a
valence-bond solid in Reference \citenum{Tamura09a}). Under a pressure
of 0.18 GPa this insulating state becomes a superconducting
\cite{Tamura09a}.  This suggests that under a small structural
modification to the material, the nearest-neighbor pairs in the PEC
state can become mobile, in a realization of the Schafroth theory of
local-pair SC \cite{Schafroth55a,Mazumdar08a}.  In this scenario, the
application of external pressure will strongly affect the the weakest
bonds in the crystal lattice. The weak bonds are also those
responsible for the frustration; hence one effect of pressure is to
{\it increase} the lattice frustration. We have proposed that
increased frustration allows fluctuations of the PEC ordered singlets,
causing the singlet pairs to have mobility.

\begin{figure}
\begin{minipage}{0.5\columnwidth}
\includegraphics[width=\columnwidth,keepaspectratio=true]{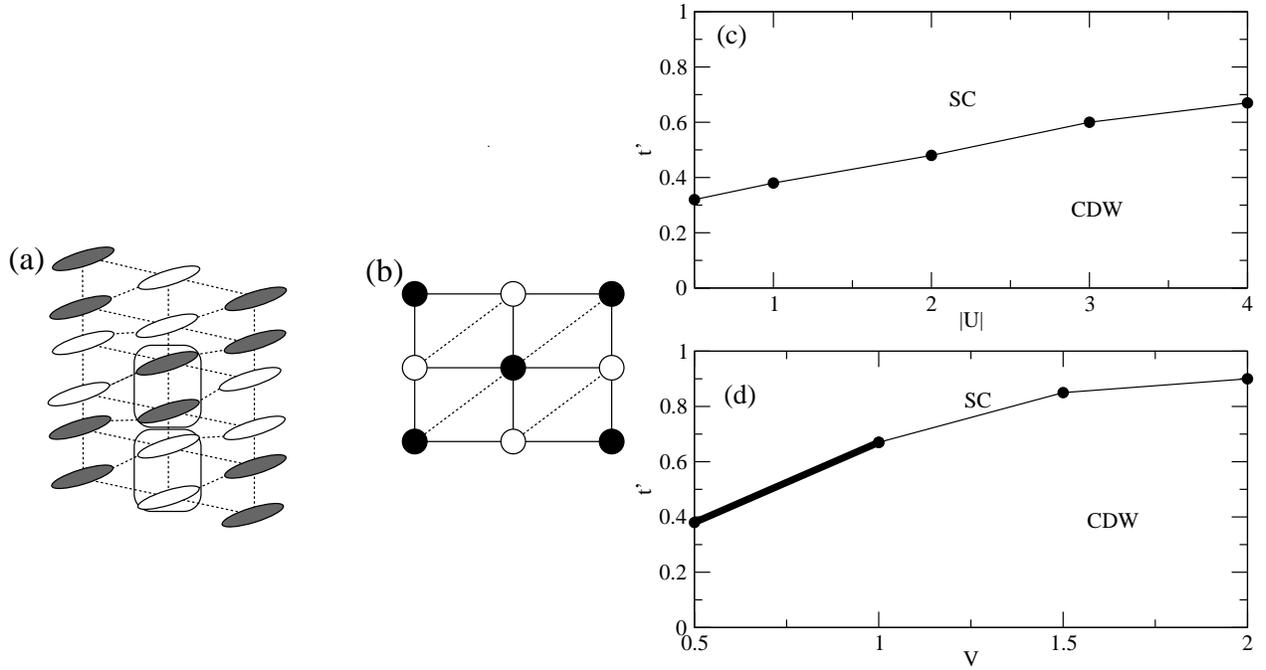}
\end{minipage}
\begin{minipage}{0.5\columnwidth}
\includegraphics[width=\columnwidth,keepaspectratio=true]{fig4cd.eps}
\end{minipage}
\caption{\label{negufig} (a) Schematic picture of the PEC insulating state
in a 2D CTS crystal. Molecules with
$\rho>0.5$ ($\rho<0.5$) are drawn with filled (open) symbols.
(b) Equivalent CO state in the effective $\rho=1$ model. Filled (open)
circles correspond to pairs of molecules with more (less) charge.
 (c) Phase diagram of the effective model (Eq.~\ref{neguham}) as a function
of $t^\prime$ and $|U|$, and (d) $t^\prime$ and $|V|$.
Reproduced from Reference \citenum{Mazumdar08a}.}
\end{figure}

A simple effective model can be constructed as shown in
Fig.~\ref{negufig}.  Fig.~\ref{negufig}(a) shows schematically the PEC
CO pattern in a 2D CTS crystal such as
EtMe$_3$P[Pd(dmit)$_2$]$_2$. Neighboring molecules with higher charge
density are singlet paired. This can be mapped to the simpler
effective model shown in Fig.~\ref{negufig}(b), where pairs of nearest neighbor
occupied (unoccupied) sites are replaced by single sites with double occupancy (vacancy). Now the CO
alternates (in the extreme limit) between charge densities of
``2'' and ``0'' carriers on each site.  This effective model therefore
has an average density of $\rho=1$ rather than $\rho=\frac{1}{2}$, and
an effective {\it attraction} between carriers on each site (negative
$U$). The long range interactions remain repulsive however. The
Hamiltonian for this model is
\begin{equation}
H=-t\sum_{\langle ij\rangle} B_{i,j} -t^\prime\sum_{[ij]} B_{i,j} -U\sum_i n_{i,\uparrow}n_{i,\downarrow} + V\sum_{\langle i,j\rangle} n_in_j.
\label{neguham}
\end{equation}
In Eq.~\ref{neguham}, operators have the same meaning as in
Eq.~\ref{ham} and Eq.~\ref{ham1d}; the important distinction is that here $\rho=1$. 
Similar modeling of spin-paired singlets by effective double
occupancies has been done in the past by others \cite{Alexandrov81a,Micnas90a}. The
difference in our work here is that the spin-paired state is not {\it
assumed} as in previous work, but is proved rigorously.

The lattice structure we chose is again a square
lattice with bonds $t$ with a frustrating bond $t^\prime$ in the $x+y$
direction.
The $-U$ interaction here leads to a superconducting phase
as expected. 
We calculated the SC pair-pair correlation function for
on-site pairs  
\begin{equation}
P({\bf r})=\frac{1}{N}\sum_j \langle c^\dagger_{j,\uparrow}c^\dagger_{j,\downarrow}c_{j+{\bf r},\downarrow}
c_{j+{\bf r},\uparrow}\rangle,
\end{equation}
and the charge structure factor
\begin{equation}
S({\bf q})=\frac{1}{N}\sum_{i,j}e^{i{\bf q}\cdot({\bf r}_i-{\bf r}_j)}
\langle (n_i-1)(n_j-1)\rangle,
\end{equation}
as a function of $t^\prime$ ($N$ is the number of lattice sites). For
small $t^\prime$ $S({\bf q})$ peaks at ${\bf q}=(\pi,\pi)$ consistent
with the checkerboard CO shown in Fig.~\ref{negufig}(b).  At a
critical $t^\prime$ a sudden decrease of $S(\pi,\pi)$ coincident with
an increase of $P({\bf r})$ indicates a transition from CO to SC.  We
show in Fig.~\ref{negufig}(c) and (d) the ground state phase diagram
of Eq.~\ref{neguham} from exact calculations on a 16 site lattice
\cite{Mazumdar08a}.  While this simple model does not capture details
of the SC state (the pairing is an on-site singlet), the calculated
frustration induced transition between CO and SC reproduces
qualitatively the experimentally observed transition from a spatial
broken symmetry state to SC in many CTS superconductors.

The mechanism for the proposed transition to the superconducting state has similarities with
some other proposed mechanisms. We have already mentioned the relationship between our work and
Schafroth's idea of the condensation of charged Bosons. Our work may also be considered as an
extension of the Resonating Valence Bond theory of SC \cite{Anderson87a} to the specific case of $\rho=\frac{1}{2}$. 
Finally, we mention the work by Dunne and Br\"andas \cite{Dunne13a,Dunne13b}, who have proposed
that condensation to the superconducting state can occur if the short-range component of the
Coulomb repulsion is screened and the long range component is attractive. In our case this 
nonlocal effective
attraction arises from the antiferromagnetic spin-spin correlations between neighboring sites
in the $\rho=\frac{1}{2}$ lattice. One difference between our work and that by Dunne and Br\"andas
is that in addition to the latter being derived from the large eigenvalue of the density
matrix and thus exhibiting ODLRO, 
in our case lattice frustration plays a key role in driving the superconducting transition,
while it is alternancy symmetry rather than frustration that is important in the model of
Dunne and Br\"andas. Further work is necessary to reveal the similarities and differences between
these models.

\section*{\sffamily \Large The ubiquity of unconventional $\rho=\frac{1}{2}$ superconductors}

In this section we point out the preponderance of correlated-electron
$\rho=\frac{1}{2}$ superconductors. In many cases, phenomenology
similar to that described above for the CTS is observed, for example
charge ordering with charge periodicity $\cdots1100\cdots$.  This is
despite radically different material characteristics (organic versus
inorganic and dimensionality).  Although the materials listed below
have attracted strong interest individually, until now the carrier
density itself was not considered an important variable.

\subsubsection*{\sffamily \normalsize Spinels}

Spinels are inorganic ternary compounds AB$_2$X$_4$, with the
B-cations as the active sites. LiTi$_2$O$_4$ \cite{Johnston73a},
CuRh$_2$S$_4$ and CuRh$_2$Se$_4$ \cite{Hagino95a} are the only three
spinels that have been confirmed to be superconductors. Ti$^{3.5+}$ in
LiTi$_2$O$_4$ has one d-electron per two Ti-ions; Rh$^{3.5+}$ in
CuRh$_2$S$_4$ and CuRh$_2$Se$_4$ is in its low-spin state and has one
d-hole per two Rh-ions. Further, Jahn-Teller distortion removes
t$_{2g}$ degeneracy, creating a true $\rho=\frac{1}{2}$ d-band of one
specific symmetry.  The crucial role of carrier density is
demonstrated from the large T$_c$ = 11 K in LiTi$_2$O$_4$ on the one hand,
and only short-range magnetic correlations down to 20 mK in LiV$_2$O$_4$. 
T$_c$/T$_F$ is recognized to be large in
LiTi$_2$O$_4$, and the mechanism of SC here remains controversial.
Importantly, static lattice distortions give a three dimensional (3D)
PEC with nearest-neighbor pairing in CuIr$_2$S$_4$ \cite{Radaelli02a}
and LiRh$_2$O$_4$ \cite{Okamoto08a}, which are isoelectronic with the
superconductors. This 3D PEC has the same {\it CO
  periodicity as in the CTS.}  Pressure-dependent measurements and
search for SC in the last two compounds are called for.

\subsubsection*{\sffamily \normalsize Na$_x$CoO$_2\cdot$yH$_2$O}

Layered cobaltates are unique in that $\rho$ can be varied over a wide
range by varying $x$ \cite{Foo04a}. We have shown that the
$\rho$-dependent electronic behavior of anhydrous Na$_x$CoO$_2$ can be
explained through an identical mechanism as in the CTS
\cite{Li11a}. In the hydrated superconducting Na-cobaltate with
$x=0.35$, the water enters as H$_3$O$^+$ ions, and the Co valence is
set by both Na doping and the amount of H$_3$O$^+$ . Experimental
measurements of the actual valence state of the Co atoms in the
superconducting compound find Co$^{3.5+}$, corresponding to
$\rho=\frac{1}{2}$ \cite{Barnes05a}.

\subsubsection*{\sffamily \normalsize Li$_{0.9}$Mo$_6$O$_{17}$}

 This material has attracted attention because of its unusually large
 upper critical field \cite{Mercure12a}.  Very little is currently
 known about the superconducting state of this material.  The
 Mo-valence of nearly 5.5 however requires equal admixing of 4d$^1$
 (Mo$^{5.0+}$) and 4d$^0$ (Mo$^{6.0+}$). It is conceivable that the
 large upper critical field is due to the local singlets with
 molecular dimension.

\subsubsection*{\sffamily \normalsize Metal-intercalated phenacenes}

 SC has very recently been found in metal-intercalated phenanthrene
 \cite{Wang11a}, picene \cite{Kubozono11a}, coronene
 \cite{Kubozono11a} and dibenzopentacene \cite{Xue12a}. In every case
 ``doping'' with 3 electrons per molecule is essential for SC. The lowest
 unoccupied molecular orbital (LUMO) and the next higher MO (LUMO+1)
 are unusually close in these molecules (in coronene they are
 degenerate). It has been shown that with 3 electrons added the
 electron populations of the LUMO and LUMO+1 are almost 1.5 each due
 to combined bandwidth and correlation effects, and that this strongly
 suggests that the mechanism of SC in these doped polycyclic aromatics
 and the CTS are same \cite{Dutta13a}.

\section*{\sffamily \Large CONCLUSIONS}

Correlated-electron SC continues to be a formidable problem in spite
of decades-long intensive research. SC at a particular carrier
density, as well as perceptible similarity between different families
of correlated superconductors can hardly be coincidences. We believe
that both features indicate that the physics of these materials
(antiferromagnetism, charge-ordering, SC) is strongly linked to their
chemistry (stoichiometry and carrier density). Our proposed mechanism
of SC, though far from complete, offers a single unified approach to a
wide variety of systems, and can perhaps even be extended to the more
popular cuprates and Fe-compounds, where too local singlet-formation
has been suggested by many authors. The strong role of electron-phonon
interactions, as observed in many experiments, is to be anticipated at
$\rho=\frac{1}{2}$ (see Figs.~\ref{1dfig} and \ref{pecfig}), in spite of the
large T$_c$/T$_F$. Our work provides strong motivation for focused
research for a theory of correlated-electron SC in $\rho=\frac{1}{2}$.

\subsection*{\sffamily \large ACKNOWLEDGMENTS}

The present review paper grew out of an invited talk presented by one of us (SM) at
the 8th Congress of the International Society of Theoretical Chemical Physics, held
in Budapest, Hungary, August 25-31, 2013. SM is grateful to Professor Miklos Ketesz
of Georgetown University) for organizing the session on Solid State Chemistry and
for inviting him. The work by the authors was supported by the Department of Energy
grant DE-FG02-06ER46315. SM acknowledges partial support from 
NSF Grant No. CHE-1151475.

\clearpage

\end{document}